\documentclass{aa}

\usepackage{graphicx}
\usepackage{txfonts}
\usepackage{color}

\begin{document}

   \title{Estimation of space weathering timescale on (25143) Itokawa: Implications on its rejuvenation process}

   \author{Sunho Jin
          \inst{1,2}
          \and
          Masateru Ishiguro\inst{1,2}
          }

   \institute{Department of Physics and Astronomy, Seoul National University, 1 Gwanak-ro, Gwanak-gu, Seoul 08826, Republic of Korea
   \and SNU Astronomy Research Center, Department of Physics and Astronomy, Seoul National University, 1 Gwanak-ro, Gwanak-gu, Seoul 08826, Republic of Korea\\
              \email{jin@astro.snu.ac.kr, ishiguro@snu.ac.kr}
             }

   \date{Received ; accepted}

\abstract{The space weathering timescale of near-Earth S-type asteroids has been investigated by several approaches (i.e., experiments, sample analyses, and theoretical approaches), yet there are orders of magnitude differences.} 
{We aim to examine the space weathering timescale on a near-Earth S-type asteroid, Itokawa using Hayabusa/AMICA images and further investigate the evolutional process of the asteroid.} 
{We focused on bright mottles on the boulder surfaces generated via impacts with interplanetary dust particles (IDPs). We compared the bright mottle size distribution with an IDP flux model to determine the space weathering timescale.} {As a result, we found that the space weathering timescale on Itokawa's boulder surfaces is 10$^3$ years (in the range of 10$^2$--10$^4$ years), which is consistent with the timescale of space weathering by light ions from the solar wind.} {From this result, we conclude that Itokawa's surface has been weathered shortly in 10$^3$ years but portions of the surface are exposed via seismic shaking triggered by a recent impact that created the Kamoi crater.}

               
               \titlerunning{Space weathering timescales of (25143) Itokawa}

   \maketitle
%

\section{Introduction} 
\label{sect1}

Space weathering denotes any surface modification processes that may change the optical, physical, chemical, or mineralogical properties of the surface of an airless body \citep{Clark2002}. It is caused by the solar wind ion implantation and the micrometeorite bombardment \citep{Pieters2016}. The space weathering effect has been observed on lunar rocks, meteorite samples, and asteroids observed by spacecraft and telescopes \citep{Clark2002}. Particularly, materials that consist of ordinary chondrites and S-complex asteroids indicate a decrease in albedo (i.e., darkening), reddening of the visible spectrum ($\lesssim 0.7 \ \mu$m), and shallowing of $\sim$1 $\mu$m absorption band via the space weathering \citep{Clark2002}.

Meanwhile, there is a counter-process against space weathering: rejuvenation or resurfacing, which exposes fresh materials beneath weathered surfaces. Several possible mechanisms for asteroidal resurfacing have been suggested by previous studies. First, tidal interactions with terrestrial planets would trigger resurfacing of the asteroid \citep{Binzel2010}. Seismic shaking by non-destructive impacts would induce granular convection that also rejuvenates surfaces \citep{Richardson2005}. Moreover, thermal fatigue, which is caused by diurnal temperature variations, would break boulders and cobbles on the surface and result in the exposure of fresh materials \citep{Delbo2014}. Furthermore, Yarkovsky-O'Keefe-Radzievskii-Paddack (YORP) effect accelerates the spin rate and would cause mass shedding and global resurfacing \citep{Pravec2007,Graves2018}.
 
An S-type, near-Earth asteroid, (25143) Itokawa, is one of the most evident exhibitions of space weathering and resurfacing phenomena. The unique trait of the asteroid is a large variety of albedos and spectra on its surface, found from the multi-band imaging observation by the Asteroid Multi-band Imaging Camera (AMICA) onboard the Hayabusa spacecraft \citep{Saito2006}. Previous studies proved that space weathering is the primary cause of albedo and spectral variation. \citet{Hiroi2006} investigated the Near-Infrared Spectrometer (NIRS) data onboard the Hayabusa spacecraft and constructed modeled spectra of Itokawa as a mixture of the spectrum of an LL5 chondrite (Alta'meem) and nanophase iron, taking account of the space weathering. \citet{Ishiguro2007} presented a global map of space weathering degrees using AMICA images. More recently, \citet{Koga2018} conducted a principal component analysis on multi-band spectra derived from AMICA images and confirmed that the main trend of the spectral variation is consistent with spectral alteration by laboratory simulations of the space weathering. Moreover, weathered rims found from the returned samples are regarded as the most definitive evidence for the occurrence of space weathering on the asteroid surface \citep{Noguchi2014}.

It is, however, important to note that the exposure time of the Itokawa's surface material is not well determined, although the Hayabusa project comprehensively explored the asteroid via remote-sensing observations and laboratory analyses of the returned samples. There is a large discrepancy in the estimate of the surface age up to four orders of magnitude (from 100 years to 10$^6$ years, \citealt{Bonal2015,Koga2018,Noguchi2011,Keller2014,Matsumoto2018,Nagao2011}). In addition, there is still an enormous discrepancy between mechanisms for determining the space weathering timescale of an S-type asteroid. It thus depends on the physical processes that cause the space weathering ($10^8$ years for micrometeorite impacts, \citealt{Sasaki2001}; $10^4-10^6$ years for heavy-ion irradiation, \citealt{Brunetto2006}; $10^3-10^4$ years for H$^+$ and He$^+$ ion irradiation, \citealt{Hapke2001} and \citealt{Loeffler2009}). These discrepancies are major obstacles to understanding the evolutional history of Itokawa's surface.

We propose a novel idea to estimate the Itokawa's surface age, focusing on bright mottles on the boulder surfaces to alleviate these discrepancies. It was reported that the bright mottles consist of fresh material under the weathered patina of boulders that are exposed by impacts with mm- to cm-sized interplanetary dust particles \citep[IDPs,][]{Takeuchi2009, Takeuchi2010}. Because these mottles obscure via space weathering to make them darker and redder again, the number of observable mottles is controlled by the balance of the timescale of space weathering and the IDPs impact frequency. 
We calculated the occurrence frequency of the bright mottles as a function of size and compared the frequency to the number of the bright mottles to determine the space weathering timescale on Itokawa. Here, we defined the space weathering timescale as the characteristic time needed for changing from fresh ordinary chondrite (OC)-like optical property to the typical (i.e., matured) optical property of the Itokawa surface. We describe our method in Sect. \ref{sect2} and findings in Sect. \ref{sect3}. Based on these results, we discuss the possible resurfacing mechanism which results in the large-scale optical heterogeneity in Sect. \ref{sect4}. 

    \begin{table*}
    \caption{Images used in this study}             
    \label{table1}      
    \centering                          
    \begin{tabular}{c c c c c}        
    \hline\hline                 
    File name & Filter & Date and Time (UT) & Spacecraft distance (m) & Pixel scale (mm pixel$^{-1}$) \\    
    \hline                        
    ST\_2544540977 & v & 2005-11-12 05:35:37 & 110.9 & 11.0 \\      
    ST\_2544579522 & v & 2005-11-12 05:55:52 & 59.9  & 5.9 \\
    ST\_2544617921 & v & 2005-11-12 06:05:55 & 77.9  & 7.7 \\
    ST\_2563511720 & v & 2005-11-19 20:23:36 & 80.9  & 8.0 \\
    ST\_2572745988 & v & 2005-11-19 20:26:36 & 62.9  & 6.2 \\

    \hline                                   
    \end{tabular}
    \end{table*}

    \begin{figure}
    \centering
    \includegraphics[width=\hsize]{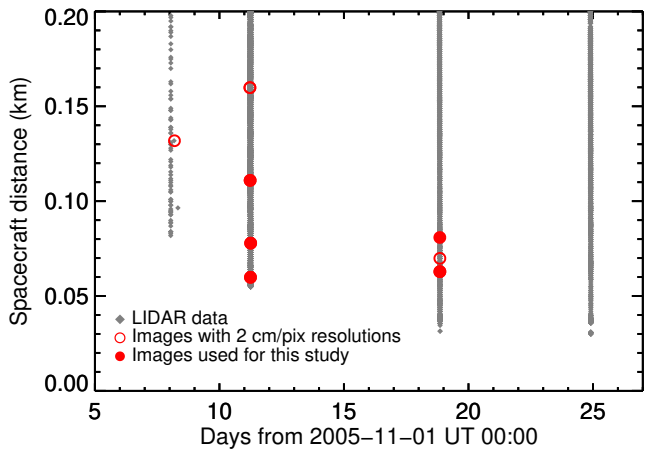}
        \caption{Distance between the spacecraft and the Itokawa's surface. The grey circles denote the data taken by Hayabusa/LIDAR in 2005 November \citep{Mukai2012}. Red (open and filled) circles indicate the distances of the spacecraft when each image was taken at distances closer than 200 m. Filled red circles show the images examined in this study.
              }
        \label{Fig1}
   \end{figure}

    \begin{figure*}
    \centering
    \includegraphics[width=\hsize]{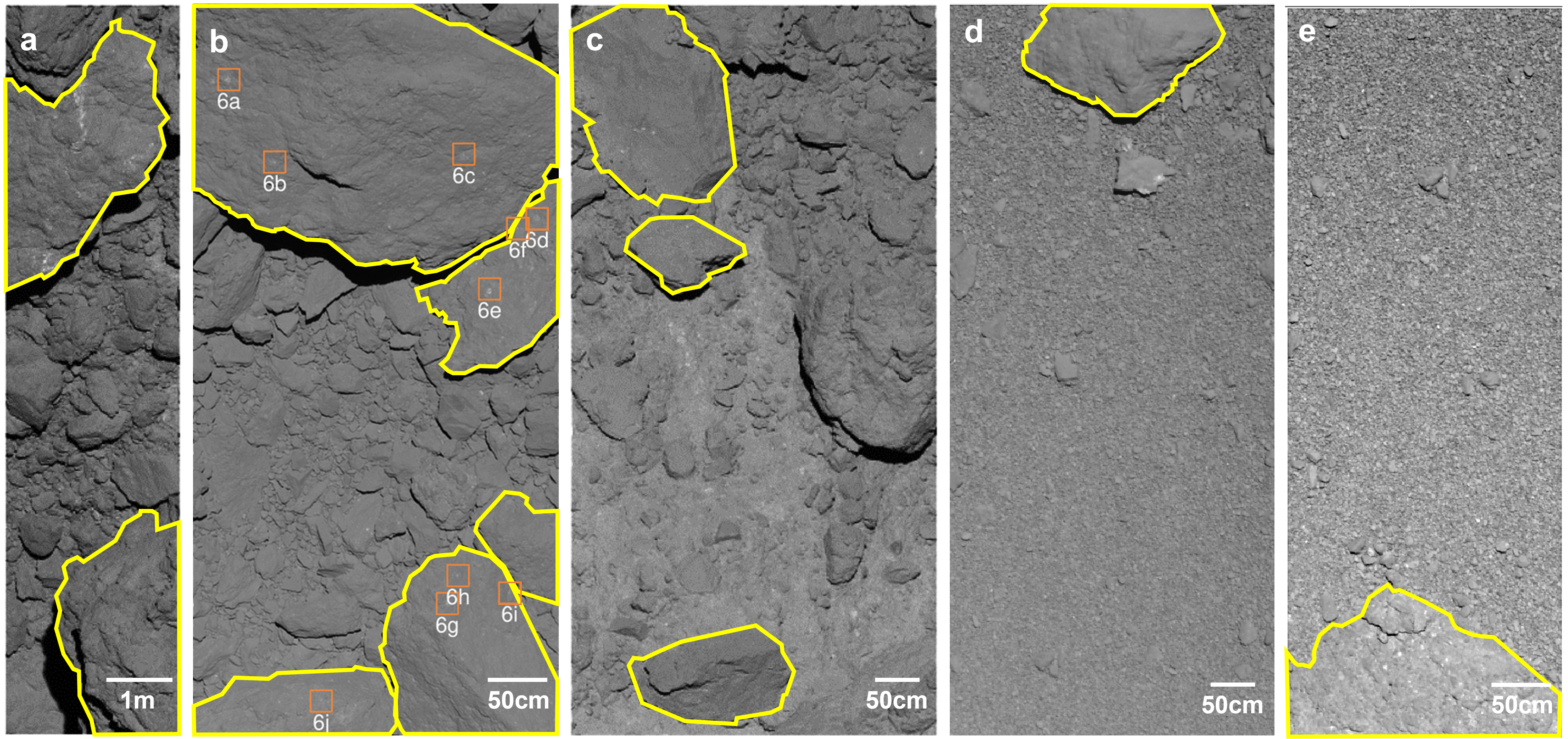}
        \caption{Five images analyzed in this study. The file names are (a) ST\_2544540977\_v, (b) ST\_2544579522\_v, (c) ST\_2544617921\_v, (d) ST\_2563511720, and (e) ST\_2572745988. We selected twelve large boulders (enclosed by yellow lines) for the analysis. We show enlarged images of the areas surrounded by orange squares in Fig. \ref{Fig6}.
              }
        \label{Fig2}
   \end{figure*}

\section{Methods}
\label{sect2}

In this chapter, we describe data preparation, the bright mottle detection technique, and a model for comparing our observational results with IDP impact flux, as shown below.


\subsection{Data preparation}
\label{sect2_1}
The Hayabusa spacecraft arrived at the gate position (about the 20 km distance from Itokawa) on 2005 September 12 and shifted to the home position (about the 7 km distance from Itokawa) \citep{Fujiwara2006}. During these phases, the mission team investigated the global structures of the asteroid using onboard instruments. In October, the spacecraft moved to several positions with different solar phase angles and approached closer distances for detailed investigations. The mission team conducted two touchdown rehearsals on 2005 November 4 and 12 \citep{Yano2006}. Finally, the spacecraft landed on the Itokawa surface on 2005 November 19 \citep{Fujiwara2006}. Figure \ref{Fig1} shows the altitudes of the spacecraft in November. This data was taken by Light Detection and Ranging instrument (LIDAR) \citep{AbeS2006,Mukai2007}. In Fig. \ref{Fig1}, we emphasized the altitude at which the AMICA images were obtained with different symbols (the open and filled red circles).

Among the imaging data available at the official website of Data Archives and Transmission System (DARTS), Institute of Space and Astronautical Science (ISAS), Japan Aerospace Exploration Agency (JAXA)\footnote{https://data.darts.isas.jaxa.jp/pub/hayabusa/} , we selected five AMICA images (ST\_2544540977\_v, ST\_2544579522\_v, ST\_2544617921\_v, ST\_2563511720\_v, and ST\_2572745988\_v) taken on 2005 November 12 and 19 (Fig. \ref{Fig2}). These images were taken during the second rehearsal and touchdown. We selected these images because they have good spatial resolutions ($<$15 mm pixel$^{-1}$ ) and contain large boulders (whose longest axis is longer than 1 m). The resolution and boulder sizes are important factors in detecting small mottles and increasing the reliability of the statistical analysis by the law of large population. We did not use ST\_2563537820\_v (an open circle at November 19 in Fig. 1) for our analysis because there is no large ($>$ 1 m) boulders in the image despite high resolution (6.9 mm pixel$^{-1}$). Detailed information on the images for our analysis is shown in Table \ref{table1}.
   
We subtracted bias and corrected flat from the raw images following \citet{Ishiguro2010}. After the preprocessing, we applied the Lucy-Richardson deconvolution algorithm to improve the blurred resolution of the AMICA images \citep{Richardson1972,Lucy1974}. The usability of the deconvolution technique is confirmed in \citet{Ishiguro2010}.

\subsection{Detection of bright mottles from images}
\label{sect2_2}
Because small boulders tend to be covered by movable regolith particles, bare rock surfaces may not be exposed on the small boulders. For this reason, we selected a total of 12 large boulders (Fig. \ref{Fig2}). These boulders have the longest axis larger than 1 m. Assuming that boulders' surfaces are perpendicular to the AMICA boresight vector, the total surface area is estimated to be 27.1 m$^2$.

\begin{figure*}
    \centering
    \includegraphics[width=\hsize]{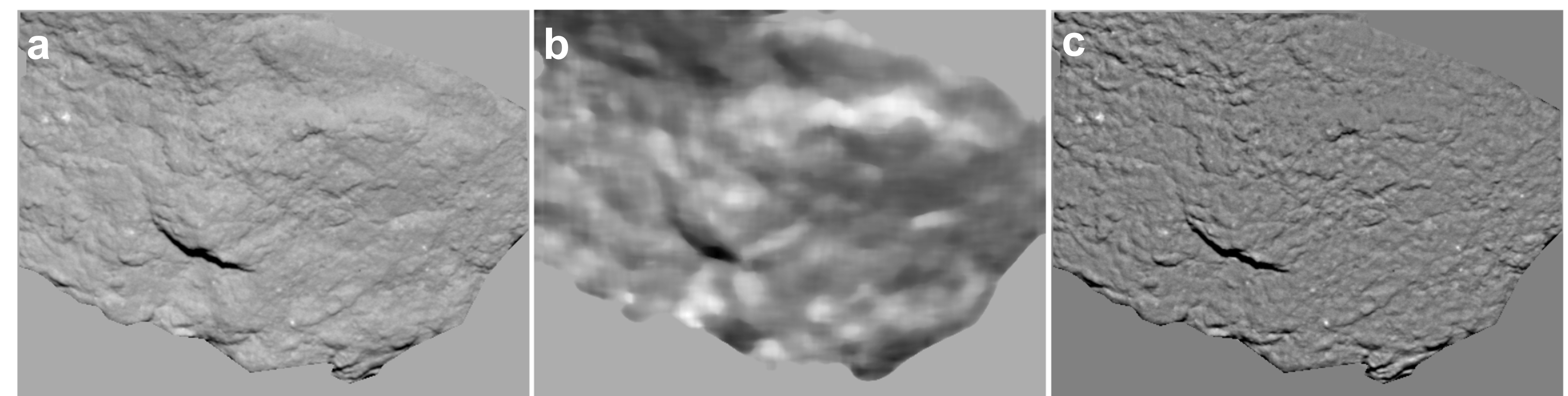}
        \caption{An example of the  (a) original, (b) median-filtered, and (c) background-subtracted images used for the analysis. The original image is a part of ST\_2544579522\_v.fits. 
              }
        \label{Fig3}
   \end{figure*}

We utilized \texttt{Source-Extractor}\footnote{https://sextractor.readthedocs.io/}, \citep{Bertin1996}) to detect bright mottles from boulders. Note that there is large scale brightness fluctuation on a boulder surface due to the different illumination conditions. This inhomogeneity is not common in astronomical images, for which \texttt{Source-Extractor} is mainly designed. Therefore, we flattened the background by subtracting smoothed images made from a 2-dimensional median filter (without using the background detection algorithm in \texttt{Source-Extractor}). We applied the 2-dimensional median filter with a square width of 19 pixels to the original image. We decided the filter size to flatten the large-scale background ($\gtrsim$10 cm) while leaving small structures of bright mottles ($\lesssim$10 cm).
Figure \ref{Fig3} a, b, c are the example of the original, median-filtered, and background-subtracted images of a boulder, respectively.


Then, we ran \texttt{Source-Extractor} with a 3-sigma detection threshold. This threshold was chosen to discriminate bright mottles from the small-scale fluctuations caused by the Poisson noise and textures of the boulders. On the other hand, the background mesh size for the calculation of background standard deviation is also 19 pixels, the same size as the median filter. The minimum area for the detection is 2 pixels to avoid false detection due to hot pixels. With this setting, we detected 499 bright sources out of 12 boulders. After this detection process, we rejected sources with elongations (the ratio of major to minor axis) larger than 2.5. This criteria is based on \citep{Elbeshausen2013}, which showed the elongation of crater is lower than 2.5 except extreme impact conditions (impact angle < 5 degrees). From this criterion, we filtered out 57 sources (11.4 percent of detected sources) and determined 442 sources as bright mottles created by impacts with interplanetary dust particles. 

We counted the number of pixels above the threshold for each bright mottle and calculated the area covered by these pixels. After that, we converted the area to the diameter of a circle with an equivalent area. Hereafter, we refer to this diameter as the size of the bright mottle. Once we obtained the size, we derived cumulative size-frequency distributions (SFDs) of bright mottles on ten boulders. We employed a logarithmic bin size \citep{Crater1979}. The range and bin size of the crater's SFDs are given in Table \ref{table2}.

\begin{table*}
\caption{Notations and ranges of discrete values for impactor's mass, velocity, mass density, and crater diameter used in Sect.  \ref{sect2_3}} 
\label{table2} 
\centering          
\begin{tabular}{lccrrrr} 
\hline\hline
           
 & Mean$^{(a)}$ & Notation$^{b)}$ & Min$^{(c)}$ & Max$^{(d)}$ & Width$^{(e)}$ & Num$^{(f)}$\\
\hline                    
Crater diameter (mm) & $\bar{D}$ & $D_l$ &  2.6 &  2.6$\times$10$^4$ & 2$^{0.25} D_{l,\textrm{min}}$  & 50\\
Impactor mass (g) & $\bar{m}$ & $m_i$ & 1.1$\times$10$^{-6}$  & 8.9 & 10$^{0.1} m_{i,\textrm{min}}$  &   69 \\
Impactor velocity (km s$^{-1}$) & $\bar{v}$ & $v_j$ & 0.5 &  89.5  & 1.0 & 90\\
Impactor mass density (g cm$^{-3}$) & $\bar{\delta}$ & $\delta_k$ &  0.125 & 7.975 & 0.05 &  158 \\
\hline                    
\end{tabular}
		\tablefoot{		                     
			\tablefoottext{a}{Mean value in each bin,}
			\tablefoottext{b}{Notation of the quantity,}
			\tablefoottext{c}{Minimum value,}
			\tablefoottext{d}{Maximum value,}
			\tablefoottext{e}{Bin width,}
			\tablefoottext{f}{and the number of bin, where \textit{i} and \textit{l} are the ordinal numbers of each bin.}}
\end{table*}


 \subsection{IDP impact model}
 \label{sect2_3}
As described above, we consider that recent IDP impacts on the bare boulder surface formed bright mottles. Accordingly, if the IDP impact flux is known, it is possible to derive the number of mottles and compare it to the numbers of the detected bright mottles. We utilized the Meteoroid Engineering Model Version 3 (MEM3, \citealt{Moorhead2020}) model to derive the IDPs impact flux colliding with boulder surfaces. This model was developed for the risk assessment of spacecraft navigating in the near-Earth region (the heliocentric distance between 0.2 au and 2.0 au). It is also applicable to any celestial bodies if the orbital information is given. We obtained Itokawa's orbital information from the JPL Horizons Web interface \footnote{https://ssd.jpl.nasa.gov/horizons.cgi}. This ephemeris includes state vectors of Itokawa with respect to Earth, starting from 2019 June 10 (JD 2\,458\,644.5) to 2020 December 16 (JD 2\,459\,199.5) for 555 days (approximately one orbital period of Itokawa, \citealt{Fujiwara2006}). We assumed that the flux averaged over 1 orbital period remained as a constant since the orbit of the asteroid has not been significantly altered during 1 Million years \citep{2002ESASP.500..331Y}.  Because the rotation axis of Itokawa is nearly aligned to the ecliptic south pole ([$\lambda,\beta$]=[128\fdg5, -89\fdg66], where $\lambda$ and $\beta$ are ecliptic longitude and latitude of the pole orientation, \citealt{Demura2006} and \citealt{Fujiwara2006}), in addition, the boulders for our analysis distribute near the equatorial region, we employed azimuthally-averaged flux. It is the impact flux to a target body rotating around the ecliptic pole and averaged along the azimuth direction. 

\begin{figure}
    \centering
    \includegraphics[width=\hsize]{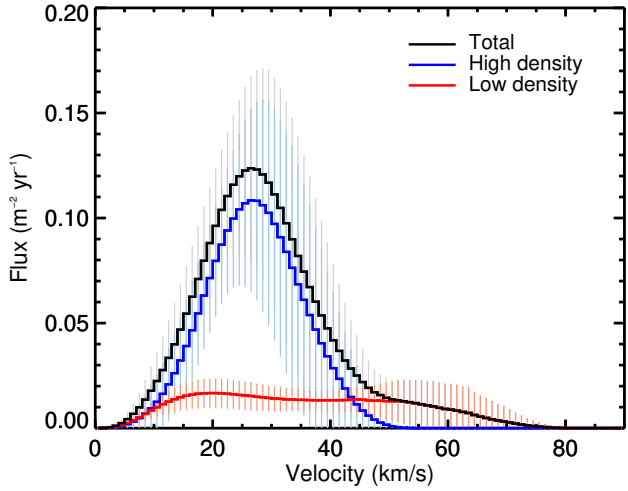}
        \caption{Cumulative IDPs flux averaged over Itokawa's one orbital revolution around the Sun. 
        The error bars correspond to the range of the IDP flux during the one orbit revolution. 
              }
        \label{Fig4}
   \end{figure}

   The MEM3 model assumes two meteoroid populations, namely, high and low-density populations with different mass densities based on \citet{Kikwaya2011}. For each population, the MEM3 model calculates the impact flux per square meter per year, $H(v)$, as a function of the impactor's velocity $v$ in the mass range of $m_\mathrm{min} \leqq m \leqq m_\mathrm{max}$, where $m_\mathrm{min}=10^{-6}$ g and $m_\mathrm{max}=10$ g are given, respectively. In Fig. \ref{Fig4}, we show  $H(v)$ for a target in the Itokawa's orbit, where we specified the velocity interval of $\Delta v=1$ km s$^{-1}$ for the calculation.
   
$H(v)$ is written as
   
    \begin{equation}
       H(v)=\int_{m_\mathrm{min}}^{m_\mathrm{max}} f(m,v)\,dm ~,
   \label{eq1}
   \end{equation}
   where $f(m,v)$ is a differential impact flux distribution with respect to $v$ and $m$, per square meter per year. It is important to note that the mass dependency of the impact flux is not available in $H(v)$ because it has an integrated form with respect to $m$. Accordingly, $f(m,v)$ in Eq. (\ref{eq1}) is more useful than $H(v)$ for our study because we need to compare the size (derivable from $m$) frequency distribution of the bright mottles with a model. Following the recommendation in \citet{Moorhead2020}, we incorporated the cumulative IDPs flux model $F_{\textrm{Gr\"{u}n}}(m)$ in \citet{Gruen1985} into $H(v)$ obtained by the MEM3 model. It is given by
   
   \begin{equation}
            F_{\textrm{Gr\"{u}n}}\left(m\right)
            =\left(c_1 m^{\gamma_1}+c_2\right)^{\gamma_2} 
            +c_3 \left(m + c_4 m^2 + c_5 m^4 \right)^{\gamma_3} 
            +c_6 \left(m + c_7 m^2 \right)^{\gamma_4}~,
      \label{eq2}
   \end{equation}
   
\noindent where
   $c_1=2.2\times10^3$,
   $c_2=15$,
   $c_3=1.3\times10^{-9}$,
   $c_4=10^{11}$,
   $c_5=10^{27}$,
   $c_6=1.3\times10^{-16}$,
   $c_7=1.0\times10^6$,
   $\gamma_1=0.306$,
   $\gamma_2=-4.38$,
   $\gamma_3=-0.36$,
   and $\gamma_4=-0.85$ are constants. The mass, $m$, is in the unit of a gram in Eq. (\ref{eq2}).

With $F_{\textrm{Gr\"{u}n}}(m)$, the cumulative IDP flux with the particle mass larger than $m$ is given as a function of mass and velocity:

   \begin{equation}
       F(m, v)=H(v) \frac{F_{\textrm{Gr\"{u}n}}(m)}{F_{\textrm{Gr\"{u}n}}(m_\mathrm{min})}~,
      \label{eq3}
   \end{equation}
\noindent where we chose the denominator (the cumulative flux for $m > m_\mathrm{min}$) to conserve the total flux.  Because the MEM3 model generates the flux for discrete velocity and mass density (see below) values, we hereafter notate discrete values as $(m_i, v_j, \delta_k)$ rather than $(m, v, \delta)$ for mass, velocity, and mass density. Table \ref{table2} summarizes the notation and the range of these discrete physical quantities. For our convenience, we converted the cumulative flux $F_{\textrm{Gr\"{u}n}}(m_i)$ into the differential flux within \textit{i}-th mass bin as below:

   \begin{equation}
      \begin{split}
       f(m_i, v_j)&=F(m_i, v_j)-F(m_{i+1}, v_j) \\
       &= \frac{F(m_\mathrm{min}, v_j)}{ F_{\textrm{Gr\"{u}n}}(m_\mathrm{min})} \Bigl( F_{\textrm{Gr\"{u}n}}\left(m_i\right)- F_{\textrm{Gr\"{u}n}}\left(m_{i+1}\right)\Bigr) ~.\\
       \end{split}
      \label{eq4}
   \end{equation}
   
The MEM3 model also provides a probability distribution of the mass density $\delta_k$. The probability distribution function, $C(\delta_k)$, is defined as the ratio of the number of particles within a given density bin to the total number of particles. In the MEM3 model, $C(\delta_k)$ is independent of $m_i$ and $v_j$. With this function, the IDP flux for a given mass, velocity, and mass density is calculated from
   
   \begin{equation}
       f'(m_i, v_j, \delta_k)=f(m_i, v_j)~ C(\delta_k) ~.
      \label{eq5}
   \end{equation}
       

   Next, we considered the crater size generated by an IDP impact with a given $m_i$, $v_j$, and $\delta_k$. We utilized the crater size model in \citet{Holsapple1993}. We thus calculated the cratering volume $V(\bar{m}_i, \bar{v}_j, \bar{\delta}_k)$ excavated by an impact with given mean mass $\bar{m}_i$ ($:=(m_i+m_{i+1})/2$), mean velocity $\bar{v}_j$ ($:=(v_j+v_{j+1})/2$), and mass density $\bar{\delta}_k$ ($:=(\delta_k+\delta_{k+1}/2$) by following equations
   
   \begin{equation}
      \pi_{V}     = K_1 \left( \pi_{2} \pi_{4}^{\frac{6\nu-2-\mu}{3\mu}}     +\left(K_2 \pi_{3} \pi_{4}^{\frac{6\nu-2}{3\mu}}\right)^{\frac{2+\mu}{2}}
                           \right)^{\frac{-3\mu}{2+\mu}}~,
      \label{eq6}
   \end{equation}
   \begin{displaymath}
      \pi_{V}      = \frac{\rho V}{\bar{m_i}}    \, , \;
      \pi_{2}      = \frac{g \bar{a}_{i, k}}{\bar{U_j}^2} \, , \;
      \pi_{3}      = \frac{Y}{\rho \bar{U_j}^2} \, , \;
      \pi_{4}      = \frac{\rho}{\bar{\delta}_k}~,
   \end{displaymath}

\noindent where $\pi_{V}$ is the so-called cratering efficiency, defined as a ratio of the crater mass to the impactor mass \citep{Holsapple1993}. $\bar{a}_{i, k}$ and $\bar{U}_j$ denote the mean radius and the normal component of the mean velocity of the impactor, respectively.  We assumed an oblique impact with the most probable impact angle $\theta=45 \degr$ \citep{Gault1978}. This assumption of the oblique impact reduces the vertical impact velocity by a factor of $\sqrt{2}$ (i.e., $\bar{U}=\bar{v}/\sqrt{2}$). The constants, $Y$, $\rho$, and $g$ are the tensile strength, the bulk density, and the surface gravity of the target body. We assumed a spherical impactor whose mean radius is given as below

   \begin{equation}
       \bar{a}_{i, k}=\left(\frac{3}{4\pi}\frac{\bar{m}_i}{\bar{\delta}_k}\right)^{1/3}~.
      \label{eq7}
   \end{equation}

To obtain the crater volume $V(\bar{m}_i, \bar{v}_j, \bar{\delta}_k)$, we used Eq. (\ref{eq6})--(\ref{eq7}) for impactors with given $\bar{m}_i$, $\bar{v}_j$, and $\bar{\delta}_k$. We substituted $\rho=3.4$ g cm$^{-3}$ based on the measurement of the bulk density of Itokawa's samples \citep{Tsuchiyama2011}. The gravitational acceleration on the Itokawa surface is given as $g=8.4\times10^{-3}$ cm s$^{-2}$ \citep{Tancredi2015}. For the other parameters for characterizing the target boulders, we assumed a hard rock-type material and referred to the values in \citep{Holsapple1993, Holsapple2022}. Table \ref{table3} summarizes the applied values for the computation. 
With Eq. (\ref{eq6}), we calculated $\pi_V$ for each impactor with given $\bar{m}_i$, $\bar{v}_j$, and $\bar{\delta}_k$, and obtained the crater mass $\rho V$. The crater radii were then derived as $R=K_rV^{1/3}$, in the case of a simple bowl-shaped crater \citep{Holsapple2022}.
   
   \begin{table}
\caption{Parameters used for the evaluation of the crater's diameter}             
\label{table3}      
\centering                          
\begin{tabular}{c c c}        
\hline\hline                 
Parameter & Applied value & Reference \\    
\hline                        
   $K_1$ & 0.06 & 1 \\      
   $K_2$ & 1 & 1 \\
   $\nu$ & 0.33 & 1 \\
   $\mu$ & 0.55 & 2 \\
   $\rho$ & 3.4 (g cm$^{-3}$) & 3 \\
   $g$ & $8.4\times10^{-3}$ (cm s$^{-2}$) & 4 \\
   $Y$ & $1.44\times10^{8}$ (g cm s$^{-2}$) & 1 \\
   $K_r$ & 1.1 & 1 \\
\hline                                   
\end{tabular}
\tablebib{
(1)~\citet{Holsapple2022}; (2) \citet{Holsapple1993}; (3) \citet{Tsuchiyama2011}; (4) \citet{Tancredi2015}
}
\end{table}
   
   \citet{Holsapple2013} asserted that craters on small (sub-km sized) asteroids are expected to be spall craters. They are a kind of craters surrounded by shallow spallation features with diameters larger than 2--4 times of those of simple bowl-shaped craters. Since the depth of the space weathered rim layer found in Itokawa samples is thin enough ($<1 \mu$m, \citealt{Noguchi2014}), it is reasonable to assume that the diameters of bright mottles are equivalent to the diameter of the bowl-shaped craters. Therefore, the diameter of bright mottles including spalled region can be given as $D=2C_\mathrm{spall}R$ $(2\leq C_\mathrm{spall}\leq 4)$. We will discuss the effect of $C_\mathrm{spall}$ in Sect. \ref{sect4_2_5}.
   
   After deriving $D$, we counted the total number of the craters within given diameter bins, $N(D_l)$. For the consideration of the diameter bins, we employed a logarithmic bin size to match the bright mottle SFD from the observation, namely,
   
   \begin{equation}
      D_l=2^{\frac{l}{4}-1} ~~~(l=1,2,\dots50)~.
   \label{eq8}
   \end{equation}  
   Then, the number of craters within \textit{l}-th diameter bin, $N(D_l)$ is counted as
    
   \begin{equation}
   \begin{split}
        N(D_{l})&=\sum_{l=1}^{50} cf'(\bar{m}_i, \bar{v}_j, \bar{\delta}_k) \begin{cases}
    c=1,& \text{if } D_{l}\leq D(\bar{m}_i, \bar{v}_j, \bar{\delta}_k) \leq D_{l+1}  \\
    c=0,& \text{otherwise} ~~,
\end{cases}
    \end{split}
   \label{eq9}
   \end{equation}
   
\noindent where the subscript $l$ is an ordinal number up to 50 (i.e., $l=1,2, \dots, 50$). 

   
\section{Results}
 \label{sect3}
We identified 442 bright mottles from twelve boulders (the projected total area of 27.1 m$^2$).  The average spatial density is 16.3 m$^{-2}$. Hereafter, we show our findings as below.

\subsection{Cumulative size-frequency distributions (CSFDs)}
\label{sect3_1}

\begin{figure*}
\centering
 \includegraphics[width=\hsize]{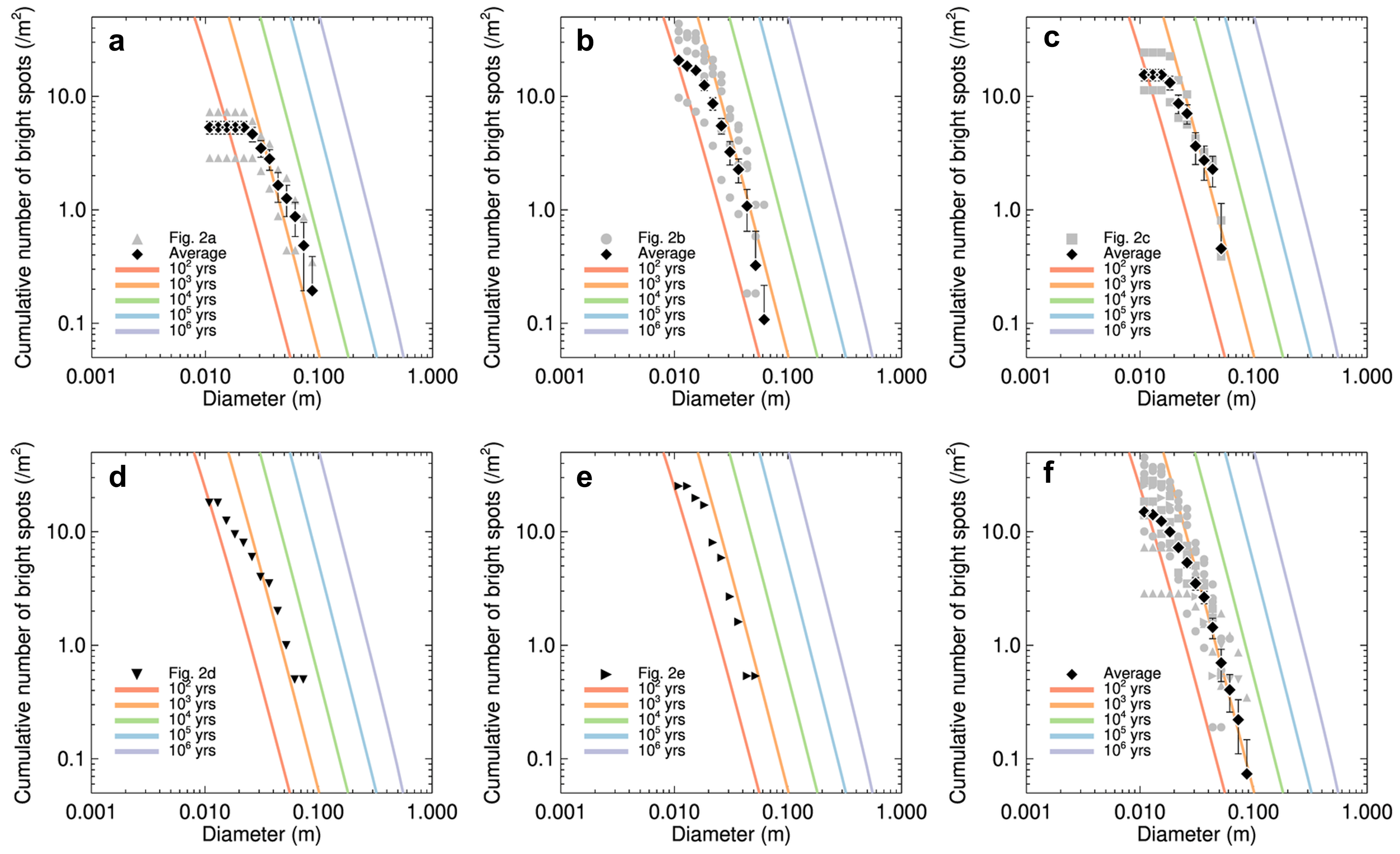}
\caption{CSFDs of bright mottles per unit area (black diamonds) compared to the estimated CSFD from the IDPs impact models (Colored lines). Error bars indicate a 1-$\sigma$ confidence interval, which assumes Poisson distribution. Panels with the labels (a), (b), (c), (d), and (e) are obtained from different images (ST\_2544540977, ST\_2544579522, ST\_2544617921, ST\_2563511720, and ST\_2572745988) and one with label (f) shows the average of 5 images listed above.
              }
\label{Fig5}
\end{figure*}

Figure \ref{Fig5} indicates the cumulative size-frequency distributions (CSFDs) of bright mottles per unit area on each boulder. Each panel in Fig. \ref{Fig5} (a)--(e) was obtained from different images (i.e., the different spatial resolutions), while the different markers in each panel are CSFDs on each boulder, and Fig. \ref{Fig5} (f) shows the average of all 12 boulders. At first glance, the slopes and absolute numbers of CSFDs match one another within an order of magnitude, regardless of the different images or different boulders. From this evidence, it is expected that the exposure times of each boulder are similar to each other. For all profiles, slopes of CSFDs for larger bright mottles are steeper than those for smaller ones. This is because the spatial resolution is insufficient to detect and measure small mottles with diameters equivalent to the pixel resolutions. The inflection points ($D\sim 3-4$ cm) in Fig. \ref{Fig5} (a) is larger than the inflection points ($D\sim 2-3$ cm) in Fig. \ref{Fig5} (b)--(c) due to the different spatial resolutions of each image, indicating that the inflection points are determined by artifacts of the observational resolutions.

Figure \ref{Fig5} also compares observed CSFDs with the IDPs impact model (see,  Sect. \ref{sect2_3}). The slopes of CSFDs are consistent between these observations and the IDP impact model. For the large bright mottles (3--7 cm) in Fig. \ref{Fig5} (f), the power index of CSFDs is $q=-3.67\pm0.23$, where the error stands for the standard deviation of CSFDs of different boulders. The observed power index is consistent with the IDPs impact model (i.e., $q=-3.61\pm0.03$) within the error ranges. In each figure, we multiplied the modeled impact flux by several different exposure times to space (10$^2$, 10$^3$, 10$^4$, 10$^5$, and 10$^6$ years). Ignoring the observed CSFDs in the small size range, we found that the observed CSFDs match the IDPs CSFDs with the exposure time of $\sim$1\,000 years. Accordingly, we conclude that impact-triggered bright mottles have been obscured by the space weathering effect that has changed the reflectances of the bright mottles as dark as the surrounding areas in a timescale of 1\,000 years.

\subsection{The morphology}
\label{sect3_2}

\begin{figure*}
\centering
\includegraphics[width=\hsize]{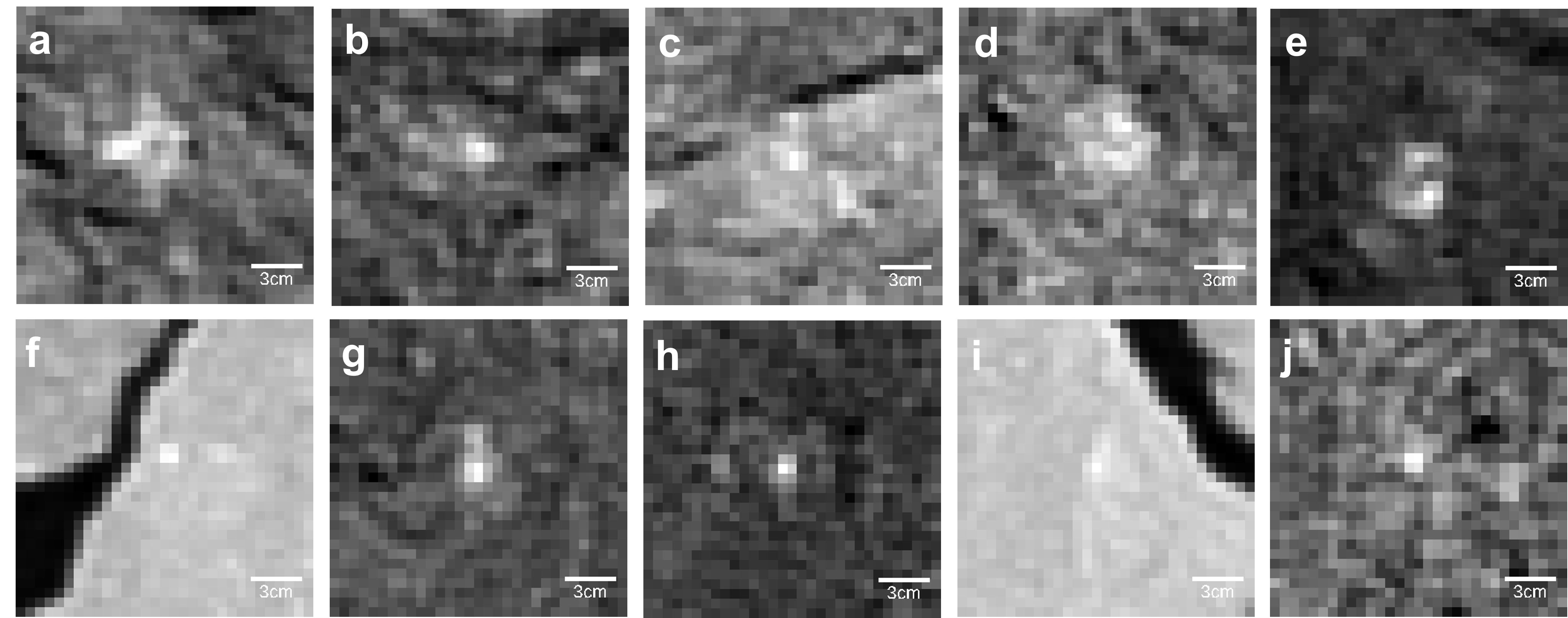}
\caption{Close-up images of the largest bright mottles. Labels in each panel correspond to the areas in Fig. \ref{Fig2}b.}
\label{Fig6}
\end{figure*}

As mentioned above, we estimated the space weathering timescale assuming that bright mottles are impact craters with pit-halo structures. The pit-halos have a rounded central pit surrounded by a region of partially excavated material. To confirm the existence of the pit-halo structures, we examined the morphology. Figure \ref{Fig6} is enlarged images of the largest bright mottles. Because the largest bright mottle (Fig. \ref{Fig6} a) has a diameter of 8 pixels (5 cm), it is possible to confirm the detailed shapes for large bright mottles. We identify at least three halo features (Fig. \ref{Fig6} a, d, and e) enclosed the central holes. The pit-halo features are unclear for the mottles smaller than $\sim$3 cm because of the insufficient image resolution. The diameter ratio between the central depression (the dark parts in the center) to the surrounding halo is $\sim$2.8 by visual inspection, which is in accordance with the general value for pit-halo craters \citep{Holsapple2013}. For this reason, all detected bright mottles are likely accompanied by pit-halo structures.

\section{Discussion}
\label{sect4}

In this section, we compare our result with the previous research in Sect. \ref{sect4_1} and discuss the uncertainties of our approach for deriving the space weathering timescale ($T_\mathrm{SW}$) in Sect. \ref{sect4_2}.  Lastly, we consider a possible phenomenon that might have occurred on the Itokawa surface in about the past $\sim$10$^3$ years based on the derived $T_\mathrm{SW}$ in Sect. \ref{sect4_3}.

\subsection{Comparison with previous research}
\label{sect4_1}

$T_\mathrm{SW}$ has been investigated by laboratory experiments and theoretical approaches. In the early study of space weathering, \citet{1999EP&S...51.1255Y} and \citet{Sasaki2001}  conducted irradiation in a space weathering simulation for the micrometeorites impacts. \citet{Sasaki2001} found that the spectrum of olivine pallet samples indicated a spectrum consistent with A-type asteroids (olivine-rich S-type) after 30-mJ irradiation of the laser. From this experiment, they derived $T_\mathrm{SW}$=10$^8$ years in case of the space weathering of dust impact heating \citep{Sasaki2001}. 

On the other hand, \citet{Loeffler2009} investigated the solar wind's influence by conducting a He$^+$ ion irradiation experiment on their olivine samples and applied their experimental result to objects at 1 au. They found that the characteristic $T_\mathrm{SW}$ induced by the solar wind He$^+$ ion irradiation is $T_\mathrm{SW}$=(5--1.3)$\times$ 10$^3$ years. \citet{Hapke2001} also estimated that the space weathering time required to alter asteroid soil by Hydrogen ions is about 5$\times$ 10$^4$  years at 3 au. Assuming that the darkening time is inversely proportional to the solar wind flux, we derived $T_\mathrm{SW}$=8$\times$ 10$^3$ years on Itokawa's orbit (the semi-major axis $a$=1.32 au and the eccentricity $e$=0.28) from this Hapke's estimate for H+ ion irradiation.

The space weathering time caused by the heavy ion irradiation was also examined. \citet{2005Icar..174...31S} investigated the spectral alteration of an ordinary chondrite (OC) meteorite (H5) by irradiating with Ar$^{2+}$ to simulate heavy ion irradiation in the solar wind and estimated $T_\mathrm{SW}=1.3\times10^6$ years for S-type near-Earth (including Itokawa). Considering elements heavier than argon, \citet{2005Icar..174...31S} further found that $T_\mathrm{SW}$ by heavy elements is on the order of $10^4$-- $10^6$ years for S-type near-Earth asteroids. \citet{2005Icar..179..265B} performed ion irradiation experiments using four different ions (H$^+$, He$^+$, Ar$^+$, and Ar$^{2+}$) and derived $T_\mathrm{SW}<10^6$ years. 

Comparing these experimental results, our observational research of bright mottles is consistent with the estimate by the light elements (He$^+$ and H$^+$) experiments and close to the lower end estimate by the heavy elements. However, it is about five orders of magnitude shorter than the estimate of dust impact heating.

Conclusive evidence of space weathering (i.e., nano-phase irons) 
 by solar wind ions was found in samples from Itokawa \citet{Noguchi2011}. In addition, noble gas elements (He, Ne, and Ar) from the Sun trapped in various depths of Itokawa's samples were also detected. \citet{Keller2014} used solar flare track density to date regolith samples to $10^2$ to $10^4$ years. \citet{Matsumoto2018} compared the size distribution of microcraters on the Itokawa sample surface and the lunar secondary impact fluxes and estimated the direct exposure timescale of Itokawa regolith particles as $10^2$ to $10^3$ years. \citet{Nagao2011} estimated the space weathering ages of Itokawa samples to be 150--550 years based on an analysis of He concentration in the samples. Our estimate of the space weathering timescale is also consistent with the exposure time of the regolith particles. 
 

Itokawa's surface age has also been determined from remote sensing observation data ($10^5$--$10^6$ years, \citealt{Bonal2015,Koga2018}). However, as already pointed out in \citet{Bonal2015} and \citet{Tatsumi2018b}, the estimate changes by one or two orders of magnitude depending on the experimental data used to convert spectra and colors to ages. Therefore, it is safe to say that our result does not contradict previous measurements using the remote-sensing data.

\subsection{Uncertainty analysis}
\label{sect4_2}
We estimated the space weathering timescale to be 1\,000 years based on the CSFDs of bright mottles compared with the IDP impact model. It is important to scrutinize problems hidden behind our analysis, the model, and assumptions and clarify the uncertainty of the estimated space weathering timescale to assess the confidence in the result. The following is a list of uncertainties related to our analysis, measurements, and assumptions:

\begin{itemize}
   \item Uncertainty in the mottle counting  
   \item Possibility of the false detection
   \item Uncertainty of the crater size measurement
   \item Uncertainty of the MEM3 IDP impact model
   \item Uncertainty for converting from crater size to impactor size.
\end{itemize}

\noindent In the following subsections (Sect. \ref{sect4_2_1}--\ref{sect4_2_5}), we discussed these uncertainties and examined the impact on our result.

\subsubsection{Uncertainty in the mottle counting}
\label{sect4_2_1}
We set a 3-sigma threshold for the mottle detection and counting. This threshold may underestimate the number of mottles because there would be fainter mottles under the detection limit. We consider how sensitive our mottle detection algorithm is for space weathering research. We made the following estimate based on the radiance factor, RADF, in Fig. \ref{Fig2}b. We converted the observed counts, $I(x, y)$, into radiance factor using the equation below:

\begin{equation}
    \textrm{RADF}=\frac{C_0~ I(x, y)}{t_{\textrm{exp}}}\frac{\pi r^2}{S_\textrm{v}}~,
    \label{eq10}
\end{equation}

\noindent where $C_0$ and $t_{\textrm{exp}}$ are the calibration factor (for v-band data, $3.42\pm0.10 \times 10^{-3}$ (W m$^{-2}$ $\mu$m$^{-1}$ sr$^{-1}$) / (DN s$^{-1}$), \citealt{Ishiguro2010}) and the exposure time, respectively. $x$ and $y$ denote the pixel coordinate on the AMICA images. $S_v$ is the solar irradiance in v-band at the heliocentric distance of $r$ in au\footnote{https://www.nrel.gov/grid/solar-resource/spectra.html}. With Eq. (\ref{eq10}), we found that the average RADF value and the 3-sigma detection threshold are 0.193 and 0.208. Our detection algorithm cannot extract mottles if they are fainter than RADF$=$0.208 (8\% excess of the ambient weathered region).


The albedo of ordinary chondrite (OC) is expected to decline precipitously in the early stages of space weathering evolution and reach a constant value when the abundance of nano-phase irons saturates in the rims of OC materials. \citet{Shestopalov2013} investigated the time evolution of albedo for OCs (including LL6, an analog of Itokawa) and suggested that the albedo dropped from 0.17--0.52 to $\approx$0.05 in the early stage and reached a nearly constant value. Although the definition of the albedo is not described explicitly in \citet{Shestopalov2013}, it seems to us that it is the Bond albedo because they compared the albedo with meteorite spectra in \citet{1976JGR....81..905G}, where the spectral data are comparable to the Bond albedo. From the low Bond albedos of Itokawa (0.02$\pm$0.01, \citealt{2008EP&S...60...49L}), most of the Itokawa surface material is likely weathered to some degree, not fresh material. Moreover, because the 3-sigma detection limit of our algorithm captures an 8\% albedo excess (significantly smaller than the initial drop in albedo), we think the detection capability of the bright mottles is high enough to characterize the initial precipitous darkening phase by the space weathering. Whereas there can be faint mottles under our detection limit, we consider that they were almost saturated by the initial space weathering effect and indicated slow albedo decrease in the matured phase.


\subsubsection{Possibility of the false detection}
\label{sect4_2_2}
We anticipate some objections to our assumption of the pit-halo crater because most of the bright mottles are not resolved in the AMICA images. Accordingly, some detected mottles may not be pit-halo craters but bright inclusions. For instance, chondrules with a large reflectance might be exposed on the surface and mistakenly recognized as pit-halo craters. However, we would argue that such a false counting of the bright inclusions is less likely because the typical chondrule size in LL type OC is $\sim$1 mm on average (up to 3.5mm), even smaller than the detected bright mottles \citep{2015ChEG...75..419F}. Moreover, the consistency in the slopes of the CSFDs between our IDP impact model and the detected mottles implies that the detected bright mottles are likely the impact origin. The pit-halo structure and quasi-circular morphology found in the close-up images of large mottles (see, Fig. \ref{Fig5}) also supports the assumption of the impact origin. For these reasons, we would assert that the influence of false detection can be negligible, especially in the large size range.

\subsubsection{Uncertainty of the crater size measurement}
\label{sect4_2_3}
The diameters of bright mottles are determined from the observed images. Thanks to the image deconvolution technique, the image resolutions are comparable to the pixel resolutions (i.e., 6--11 mm pixel$^{-1}$). The resolutions are sufficient to derive the sizes of the large mottles (D=3--5 cm, 5--8 pixels) with an accuracy of $\approx$10--20 \%. However, as we mentioned in Sect. \ref{sect3_1}, the derived diameter would be less accurate for the small mottles (the diameter $<$3 cm). In fact, the CSFDs do not match the IDP impact model in this small size range ($<$2–-3 cm), probably because of the lack of image resolutions. 

To summarize our discussion so far (Sect. \ref{sect4_2_2}--\ref{sect4_2_3}), we can assert that our estimate of the space weathering timescale is sufficiently reliable because we focused on CSFDs with a reliable size range ($>$3 cm).

\subsubsection{Uncertainty of the IDP flux model}
\label{sect4_2_4}
\citet{Moorhead2020} compared the MEM3 IDP flux model with in-situ measurements. They converted collision records from Pegasus satellites and Long Duration Exposure Facility (LDEF) into flux using ballistic limit equations. They found that the impact rate predicted by the MEM3 model is 2--3 times lower than the Pegasus measurement. On the contrary, the MEM3 model indicated the impact rate two times as high as the LDEF observation. \citet{Moorhead2020} interpreted that these discrepancies are within a predicted range because the inherent uncertainty of Gr\"{u}n's model (the underlying model for MEM3) near 1 au is a factor of $\sim$3 of the nominal flux \citep{Drolshagen2009}. Therefore, the space weathering timescale also has a factor of 3 uncertainty associated with the MEM3 IDP flux model.

\subsubsection{Crater size relation and spallation}
\label{sect4_2_5}

We employed Holsapple's scaling law for deriving the diameter of the bright mottles produced by IDP impacts (Sect. \ref{sect2_3}). Although this Holsapple's scaling law has been widely used, it is worthwhile testing differences in crater diameters using a different model.

Furthermore, Suzuki's group performed impact experiments on porous targets. \citet{suzuki2021} conducted oblique impact experiments on a porous target and found that the pit-halo structure disappeared. Although we assumed in this paper that Itokawa boulders have low porosity, it would be important to consider the possibility of boulders with high porosity. However, it is unlikely that the boulders we analyzed are as porous as \citet{suzuki2021} did in their oblique impact experiment because the pit-halo craters are found on the boulders. In addition, the low porosity of LL chondrites and the Itokawa samples (0-10 \% with an average of 1.5 and 1.9 \%, respectively) may also support our argument for low porosity \citep{Tanbakouei2019}.

Moreover, there is an ambiguity in the model associated with the diameter ratio of pit-halo craters to bowl craters. It is between 2 and 4 \citep{Holsapple2013}. We adopted only the intermediate value 3 in Sect. \ref{sect2_3}. Varying this ratio of 2 (lower limit) and 4 (upper limit), we find that the space weathering timescale changes by a factor of about 4 and 0.35, respectively.

Table \ref{table4} summarizes the major uncertainty factors and upper/lower limits of the space weathering timescales. Even with all these uncertainties, the error in our estimate of the space weathering timescale is likely to be smaller than an order of magnitude.

   \begin{table}
\caption{Upper and lower limits of space weathering timescale for each uncertain factor. $T_\textrm{sw}$ denotes the nominal value of $10^3$ years.}             
\label{table4}      
\centering                          
\begin{tabular}{c c c c}        
\hline\hline                 
Uncertainty factors & Minimum & Maximum & Reference\\    
\hline                        
    MEM3 model & 0.33 $T_{\textrm{sw}}$ & 3 $T_{\textrm{sw}}$ & 1, 2 \\      
    Cratering scaling law & 0.5 $T_{\textrm{sw}}$ & $T_{\textrm{sw}}$ & 3, 4\\
    Pit-halo diameter & 0.35 $T_{\textrm{sw}}$ & 4 $T_{\textrm{sw}}$ & 5\\
\hline                                   
\end{tabular}
\tablebib{
(1)~\citet{Moorhead2020}; (2) \citet{Drolshagen2009}; (3) \citet{Gault1973}; (4) \citet{Suzuki2012}; (5) \citet{Holsapple2013};
}
\end{table}

\subsection{Implications on resurfacing mechanism}
\label{sect4_3}

The space weathering timescale estimated in this study provides important information on the evolution of Itokawa. In particular, we focus on the ubiquity of fresh regions throughout the Itokawa surface. It is interesting to consider how fresh surfaces are exposed despite the short timescale of space weathering ($\sim$1\,000 years). Hereafter, we consider the possible mechanisms for exposing fresh surfaces.

Tidal resurfacing might have triggered a large-scale exposure of fresh materials \citep{Binzel2010}. However, \citet{Binzel2010} suggested that Q-type asteroids underwent close encounters with terrestrial planets within 5$\times10^{5}$ years, longer than the space weathering timescale derived from this study. Besides, \citet{2002ESASP.500..331Y} reported that Itokawa has maintained the present orbit for several thousand years or even much longer in the past, suggesting little chance for a planetary encounter over the space weathering time scale. Thermal fatigue caused by diurnal temperature gradient would have also created fresh materials by destructing the surface materials \citep{Delbo2014}. \citet{2018LPI....49.2628R} suggested lifetime of 10cm boulders on the surface of an S-type asteroid with a rotation period of 12 hours to be 10$^{3}$--10$^{4}$ years, comparable to the space weathering timescale of 10$^{3}$ years. This effect may be efficient near the equatorial region where the diurnal temperature variation is maximum. However, from the remote-sensing observations, fresh materials are not related to the latitude but the geological features such as steep slopes and crater rims. 

Acceleration of rotation by the YORP effect would also cause mass shedding to expose unweathered subsurfaces \citep{Pravec2007,Graves2018}. \citet{Lowry2014} found the rotation of Itokawa has been accelerated by 45 ms year$^{-1}$ during monitoring observation from 2001 to 2013. Hence it is unlikely that rejuvenation is caused by mass losses due to faster rotation in the past. Granular convection from impact-induced, global seismic shaking would expose unweathered grains on the asteroidal surface. However, this process is also unlikely to be a major resurfacing process because \citet{Shestopalov2013} show that this process only decelerates and does not counteract the space weathering. Moreover, \citet{Yamada2016} estimated that the granular convection timescale for Itokawa is the order of $10^7$ years, four orders larger than our space weathering timescale.


The remaining cause is a single impact. It is reported that the Kamoi crater (a diameter of 8 m) is the freshest terrain on Itokawa \citep{Ishiguro2007}. Assuming an oblique impact (45-degree incident angle) of an S-type impactor on Itokawa (the 0.33 km-sized S-type target asteroid), we find that a 6.5 cm-sized impactor (431 g) creates the Kamoi crater based on the Holsapple model \citep{Holsapple2022}. This size estimate is consistent with \citet{Tatsumi2018b} (20 cm) within around a factor of three uncertainty. We further estimate the impact flux on the Itokawa-sized object at 1 au is $10^{-5}$ year$^{-1}$ for this range of size/mass (6.5 cm/431 g) using Gr\"{u}n's interplanetary dust flux model \citep{Gruen1985}. Accordingly, the impact event that created the Kamoi crater is extremely rare, occurring only once every $10^{5}$ years. Although the frequency is low, we suspect that the impact that created the Kamoi crater and the subsequent rejuvenation process by seismic shaking is a possible scenario for explaining the ubiquitous exposure of fresh surfaces. The impact energy is large enough to cause global seismic shaking and induce granular convection and boulders' movements on the surface \citep{Miyamoto2014}.

Recently, \citet{Hasegawa2022} reported a spectral change of (596) Scheila as exposure of fresh surface by a large impact event in 2010. Although the spectral types are different between Itokawa (S-type) and (596) Scheila (D-type), such time-domain studies of space weathering are becoming possible. We anticipate that further experimental and observational studies of the degree of space weathering progression will be conducted.



\section{Conclusions}
\label{sect5}
In this paper, we introduced a new technique using size-frequency distributions of bright mottles at the surface of boulders to estimate the space weathering timescale of asteroid Itokawa. We suggest that the time required to alter materials with pristine state into a similar degree of the space weathering of the average Itokawa surface is $10^3$ years, with an order of magnitude uncertainty. This result is consistent with the laboratory simulation of space weathering using Hydrogen and Helium ions, the most abundant species within the solar wind. Based on our result, we conjecture that a single impact on the Kamoi crater and the subsequent seismic shaking produced the ubiquitous exposure of Itokawa's fresh surfaces.

\begin{acknowledgements}
We deeply appreciate Sasaki Sho and Emmanuel Lellouch for the valuable opinions and suggestions as a reviewer and an editor, respectively. We also thank Keith A. Holsapple and Akiko M. Nakamura (Kobe University) for helpful comments on the impact modeling of interplanetary dust particles. This work at Seoul National University was supported by the National Research Foundation of Korea (NRF) funded by the Korean government (MEST; No. 2018R1D1A1A09084105). SJ was supported by the Global Ph.D. Fellowship Program through a National Research Foundation of Korea (NRF) grant funded by the Korean Government (NRF-2019H1A2A1074796).  
\end{acknowledgements}

\bibliographystyle{aa}
\bibliography{ref_ito}
\end{document}